\documentclass[12pt]{iopart}

\usepackage{times}
\usepackage{graphicx}

\begin{document}

\title{Minimal supersymmetric SU(5) theory
and proton decay: where do we stand?}

\author{Borut Bajc 
\footnote{Talk given at Beyond the Desert 02, 
Oulu, Finland 2-7 June 2002.}}
\address{J. Stefan Institute, 1001 Ljubljana, Slovenia}

\author{Pavel Fileviez Perez}
\address{Max-Planck Institut f\" ur Physik
(Werner Heisenberg Institut), F\" ohringer Ring 6,
80805 M\" unchen, Germany}

\author{Goran Senjanovi\'c}
\address{International Centre for Theoretical Physics,
Trieste, Italy}

\begin{abstract}
We review the situation regarding $d=5$ proton decay in the
minimal su\-per\-sym\-me\-tric SU(5) GUT. The minimal theory is defined
as the theory with the minimal matter and Higgs content all
the way up to the Planck scale; of course, this allows for the
possible presence of Planck induced physics. 
It can be said that either higher dimensional operators must
be present or/and some fine-tuning  of ${\cal O}(1\%)$ of the
Higgs mass must be tolerated in order to save the theory.
\end{abstract}

\section{Introduction}

There are two popular scenarios for the solution of the
hierarchy problem. One is large extra dimensions, which
for two such new ones may be as large as a fraction of a mm 
\cite{Arkani-Hamed:1998rs,Antoniadis:1998ig,Arkani-Hamed:1998nn}. 
In this case the field-theory cutoff ($\Lambda_F$) must be
low and experiments demand: $\Lambda_F>(10-100)$ TeV. Clearly,
one then must fine-tune (somewhat) the Higgs mass, since

\begin{equation}
\label{higgs}
m_h^2\approx m_0^2+{y_t^2\over 16\pi^2}\Lambda_F^2
\approx({\rm few}\; 100\; {\rm GeV})^2\;.
\end{equation}

We believe this is acceptable; compared to the fine-tuning problem
when $\Lambda_F$ is pushed to $M_{Pl}$ (or $M_{GUT}$), this is
negligible. What is missing in this program is some serious physical
reason to have $\Lambda_F$ so low.

Another scenario is the low-energy supersymmetry, where $\Lambda_F$ gets
traded for $\Lambda_{SUSY}$ (here defined as the mass difference
between particles and superparticles of the MSSM). In this
$\Lambda_{SUSY}$ can be as low as a few hundred GeV and, strictly
speaking, no fine-tuning whatsoever is needed. On the other hand,
low-energy supersymmetry with $\Lambda_{SUSY}\approx (1-10)$ TeV 
leads to the unification of gauge couplings (predicted before experiment! 
\cite{Dimopoulos:1981yj,Ibanez:yh,Einhorn:1981sx,Marciano:1981un})
and through radiative symmetry breaking \cite{Alvarez-Gaume:1983gj} 
can explain the Higgs mechanism, i.e. the negative Higgs mass squared. 
This motivates us to focus on this scenario.

The minimal supersymmetric grand unified theory is based on SU(5)
symmetry \cite{Dimopoulos:1981zb} 
and in its minimal version contains three generations of
quarks and leptons (and their partners) and the $24$-dimensional
($\Sigma$) and $5$-dimensional ($\Phi$ and $\bar\Phi$) Higgs 
supermultiplets. In the renormalizable case there is a problem: 
one has the same Yukawa couplings for the down quarks ($Y_D$) 
and charged leptons ($Y_E$), thus

\begin{equation}
m_D=m_E\;.
\end{equation}

This works very well for $b$ and $\tau$, and fails progressively for
the second and first generation. Thus the first dilemma: to change the
theory or not? A minimalist refuses to do so, and can still invoke the 
higher dimensional operators suppressed by $\langle\Sigma\rangle/M_{Pl}$, 
where $\langle\Sigma\rangle\approx M_{GUT}$. Strictly speaking, this is 
still the minimal theory, in a sense of its structure at the scale 
$\approx M_{GUT}$, where it should be defined. In the same manner 
we speak of the standard model (SM) at $E\approx M_Z$, and the 
non-vanishing of the neutrino mass does not require changing the 
structure of the SM, but simply invoking a higher dimensional operator:

\begin{equation}
m_\nu\approx {\langle\Phi\rangle^2\over M}\;,
\end{equation}

\noindent
where $M$ is some new scale ($M\gg M_W$), for example
the mass of the right-handed neutrino in the see-saw scenario 
\cite{Mohapatra:1981yp}.

Now, supersymmetric GUTs in general, and the SU(5) theory in particular,
lead to quite fast proton decay through $d=5$ operators generated by
the superheavy coloured triplet Higgs supermultiplets ($T
$ and $\bar T$),
with masses $m_T\approx M_{GUT}$ 
\cite{Weinberg:1981wj,Sakai:1981pk} 
(more later on $m_T$). The burning issue in recent years was whether 
the minimal supersymmetric SU(5) theory is already ruled out on this basis 
\cite{Hisano:1992jj}, especially after it was found out that the RRRR 
operators play a crucial role \cite{Goto:1998qg} (in the context of SO(10) 
this was shown before in \cite{Lucas:1996bc}), and it was finally argued
last year that this was indeed true \cite{Murayama:2001ur}. 
The trouble is that one must be 
very careful in the definition of the minimal theory and it is worth
reconsidering all the issues that enter in this question and which may
save the theory. We wish to carefully discuss here all the subtleties
involved, since to us this is one of the major problems of grand
unification. After all, there is no predictive theory beyond SU(5) and
we should be absolutely sure before we rule out the only predictive
theory we have. We will see that the presence of higher-dimensional
operators and the lack of knowledge of sfermion masses and mixings may be
sufficient to make SU(5) be still in accord with experiment 
\cite{Bajc:2002bv}. If we 
require, though, absolute naturalness (no fine-tuning at all) and a
complete desert between $M_Z$ and $M_{GUT}$, the theory is ruled out.
We feel however that the above assumptions are too drastic and do not
allow for the probe of the principle of unification.

We certainly stick
to the requirement of minimality allowing for no change of the theory
all the way to the Planck scale (string scale, ...). Also, we make no
assumptions of the Yukawa couplings of the heavy particles in $\Sigma$.
Specifically, we allow for the colour octet and weak triplet in $\Sigma$
to have arbitrary masses, since the theory cannot predict them. We discuss
this below in detail.

In short, here we review the predictions of the minimal supersymmetric
SU(5) theory. We allow for arbitrary sfermion masses and mixings, keeping
of course flavour violation in accord with experiment, and we allow for small
($\approx 1\%$) amount of fine-tuning. We focus mainly on the issue of
proton decay while requiring as many as possible superpartners detectable
at LHC. It is this requirement rather than the extreme naturalness that
should make low-energy supersymmetry interesting to experimentalists
and phenomenologists (at least in our opinion). With this in mind, our
conclusion is that the minimal SU(5) theory is still in accord with
experiment, but the situation is quite tight.

Let us now systematically discuss all the issues involved in
predicting the $d=5$ proton decay amplitude:

(i) the determination of the GUT scale and the masses ($m_T$) of
the heavy triplets $T$ and $\bar T$ responsible for $d=5$ proton
decay. Specifically, we allow for arbitrary triplic couplings of the
heavy fields in $\Sigma$ and use higher dimensional terms as a
possible source of their masses \cite{Bachas:1995yt,Chkareuli:1998wi}. 
It will turn out that $m_T$ may go up 
naturally by a factor of thirty, which would increase the proton lifetime
by a factor of $10^3$.

(ii) The impact of higher dimensional operators on fermion masses and
the couplings of $T$ and $\bar T$ with fermionic supermultiplets 
\cite{Dvali:1992hc,Nath:1996qs,Nath:ft,Berezhiani:1998hg}. If
we keep the minimal theory intact, this is a must, since without
these operators fermion masses cannot be reproduced.

(iii) The freedom in sfermion masses and sfermion and fermion mixings. 
Here one must be quite careful, though, in keeping flavour violation in 
neutral currents (FCNC) in accord with experiments (for a review see for 
example \cite{Misiak:1997ei}).

Now, for generic values of the parameters of the theory we will see
that SU(5) would be ruled out. But this, however, can be said also
of FCNC in low-energy supersymmetry with generic soft terms. Instead,
we should let experiment decide the values of the parameters of the
theory. In short, with arbitrary parameters we will see that because 
of (i) to (iii) SU(5) theory is not ruled out yet.

\section{${\bf M_{GUT}}$ and ${\bf m_T}$: uncertainties}
\label{i}

The superpotential for the heavy sector is (up to terms $1/M_{Pl}$)

\begin{equation}
W=mTr\Sigma^2+\lambda Tr\Sigma^3+a{(Tr\Sigma^2)^2\over M_{Pl}}+
b{Tr\Sigma^4\over M_{Pl}}\;.
\end{equation}

Of course, if $\lambda\approx{\cal O}(1)$, we ignore higher-dimensional
terms. However, in su\-per\-sym\-me\-try $\lambda$ is a Yukawa-type coupling,
i.e. self-renormalizable. For small $\lambda$ ($\lambda\ll M_{GUT}/M_{Pl}$),
the opposite becomes true and $a$ and $b$ terms dominate. In this case,
it is a simple exercise to show that 

\begin{equation}
\label{m34m8}
m_3=4m_8\;,
\end{equation}

\noindent
where $m_3$ and $m_8$ are the masses of the weak triplet and color
octet in $\Sigma$. In the renormalizable case $m_3=m_8$.

At the one loop level, the RGE's for the gauge couplings are

\begin{eqnarray}
\label{alfa1}
\alpha_1^{-1}(M_Z)&=&\alpha_U^{-1}+{1\over 2\pi}\left(
-{5\over 2}\ln{\Lambda_{SUSY}\over M_Z}
+{33\over 5}\ln{M_{GUT}\over M_Z}
+{2\over 5}\ln{M_{GUT}\over m_T}\right)\;,\\
\alpha_2^{-1}(M_Z)&=&\alpha_U^{-1}+{1\over 2\pi}\left(
-{25\over 6}\ln{\Lambda_{SUSY}\over M_Z}
+\ln{M_{GUT}\over M_Z}
+2\ln{M_{GUT}\over m_3}\right)\;,\\
\label{alfa3}
\alpha_3^{-1}(M_Z)&=&\alpha_U^{-1}+{1\over 2\pi}\left(
-4\ln{\Lambda_{SUSY}\over M_Z}
-3\ln{m_8\over M_Z}
+\ln{M_{GUT}\over m_T}\right)\;.
\end{eqnarray}

\noindent
Here we take for simplicity $M_{GUT}=M_{X,Y}=$ superheavy gauge 
bosons masses, while at the one-loop level we could as well take 
$\Lambda_{SUSY}=M_Z$. From (\ref{alfa1})-(\ref{alfa3}) we obtain

\begin{eqnarray}
2\pi\left(3\alpha_2^{-1}-2\alpha_3^{-1}-\alpha_1^{-1}\right)&=&
-2\ln{\Lambda_{SUSY}\over M_Z}
+{12\over 5}\ln{m_T\over M_Z}
+6\ln{m_8\over m_3}\;,\\
2\pi\left(5\alpha_1^{-1}-3\alpha_2^{-1}-2\alpha_3^{-1}\right)&=&
8\ln{\Lambda_{SUSY}\over M_Z}
+36\ln{(\sqrt{m_3m_8}M_{GUT}^2)^{1/3}\over M_Z}\;.
\end{eqnarray}

This gives

\begin{eqnarray}
m_T&=&m_T^0\left({m_3\over m_8}\right)^{5/2}\;,\\
M_{GUT}&=&M_{GUT}^0\left({M_{GUT}^0\over 2m_8}\right)^{1/2}\;.
\end{eqnarray}

Since, in the case (\ref{m34m8}) is valid, $m_8\approx M_{GUT}^2/M_{Pl}$, 
we can also write

\begin{equation}
M_{GUT}\approx \left[\left(M_{GUT}^0\right)^3M_{Pl}\right]^{1/4}\;.
\end{equation}

In the above equations the superscript $^0$ denotes the values in
the case $m_3=m_8$. From (\ref{m34m8}) we get

\begin{equation}
\label{ilia}
m_T=32m_T^0\;\;\;,\;\;\;M_{GUT}\approx 10 M_{GUT}^0\;\;\;.
\end{equation}

Now, $M_{GUT}^0\approx 10^{16}$ GeV and it was shown last year 
\cite{Murayama:2001ur} that $m_T>7\times 10^{16}$ GeV is sufficiently 
large to be in accord with the newest data on proton decay. 
On the other hand, since 

\begin{equation}
m_T^0<3\times 10^{15}{\rm GeV}\;,
\end{equation}

\noindent
from (\ref{ilia}) we see that $m_3=4m_8$ is enough to save the
theory. Obviously, an improvement of the measurement of $\tau_p$
is badly needed. It is noteworthy that in this case the usual
$d=6$ proton decay becomes out of reach: $\tau_p(d=6)>10^{38}$ yrs.

\section{Higher dimensional operators and fermion masses and mixings}

In the minimal SU(5) theory at the renormalizable level we have the 
Yukawa coupling relations at $M_{GUT}$

\begin{equation}
\label{ymin}
Y_U=Y_U^T\;\;\;,\;\;\;Y_E=Y_D\;\;\;,
\end{equation}

\noindent
where in the supersymmetric standard model language the Yukawa 
sector can be written as 

\begin{eqnarray}
W_Y=&&HQ^TY_Uu^c+\bar HQ^TY_Dd^c+\bar He^{cT}Y_EL\nonumber\\
&+&{1\over 2}TQ^T\underline AQ+Tu^{cT}\underline Be^c+
\bar TQ^T\underline CL+\bar Tu^{cT}\underline Dd^c\;.
\end{eqnarray}

Also, in the minimal renormalizable model (at $M_{GUT}$)

\begin{equation}
\label{abcdmin}
\underline A=\underline B=Y_U=Y_U^T\;\;\;,\;\;\;
\underline C=\underline D=Y_D=Y_E\;\;\;.
\end{equation}

The fact that $\underline A=\underline B=Y_U$, 
$\underline C=Y_E$, $\underline D=Y_D$, is simply 
a statement of SU(5) symmetry. On the other hand 
$Y_U=Y_U^T$ and $Y_D=Y_E$ result from the SU(4) Pati-Salam 
like symmetry left unbroken by $\langle H\rangle$ and 
$\langle\bar H\rangle$. Under this symmetry $d^c\leftrightarrow e$, 
$u\leftrightarrow u^c$, $d\leftrightarrow e^c$. Of course, this symmetry is broken 
by $\langle\Sigma_\alpha^\alpha\rangle\ne\langle\Sigma_4^4\rangle$, 
where $\alpha=1,2,3$; this becomes relevant when we include 
higher dimensional operators suppressed by $\langle\Sigma\rangle/M_{Pl}$. 
The Yukawa couplings are readily diagonalized through

\begin{equation}
\label{defx}
U^TY_UU_c=Y_U^d\;\;\;,\;\;\;
D^TY_DD_c=Y_D^d\;\;\;,\;\;\;
E_c^TY_EE=Y_E^d\;\;\;,
\end{equation}

\noindent
where $X$ ($X_c$) are unitary matrices that rotate $x$ ($x^c$) 
fermions from the flavour to the mass basis and $Y_X^d$ stand for 
the diagonal Yukawa matrices. Similarly, unitary 
matrices $\tilde X$ ($\tilde X_c$) diagonalize $\tilde x$ ($\tilde x^c$) 
sfermions. The only low energy information we have is 

\begin{equation}
\label{vckm}
U^\dagger D=V_{CKM}\;\;\;,\;\;\;N^\dagger E=V_l\;\;\;,
\end{equation}

\noindent
where $V_l$ is the lepton analog of $V_{CKM}$ ($N$ rotates 
left-handed neutrinos). From (\ref{ymin}) we have 

\begin{equation}
\label{xmin}
U_c=U\;\;\;,\;\;\;E_c=D\;\;\;,\;\;\;E=D_c\;\;\;,
\end{equation}

\noindent
in the minimal renormalizable model at $M_{GUT}$.

The minimal renormalizable theory, 
as is well known, fails badly. Relations 
$m_s=m_\mu$, $m_d=m_e$ at the GUT scale are simply wrong, while 
at the same time $m_b=m_\tau$ can be considered a great success of 
the theory. We can imagine many ways out, but the simplest and the 
most suggestive is to include $1/M_{Pl}$ suppressed operators which 
are likely to be present; after all, these operators should be more 
important for the first two generations where the theory fails, and 
they require no change in the structure of the theory. This is 
analogous to a long ago speculated possibility that in the SM the 
neutrino mass is not zero, but of order $1/M$, where $M$ would 
correspond to some new physics. 

The idea of higher dimensional operators has been pursued in the 
past and applied to the proton decay issue in the case of specific mass 
textures, but never in a systematic manner. The main point is that when 
one corrects the relations (\ref{ymin}), one will also affect the 
heavy triplet couplings $\underline A$, $\underline B$, $\underline C$ 
and $\underline D$, i.e. modify (\ref{abcdmin}) which is the source 
of fast proton decay in SU(5). 

To see this, let us consider for example all the relevant 
couplings up to order $1/M_{Pl}$ ($i,j,k,l,m,n$ are SU(5) 
indices, $a,b=1,2,3$ are generation indices):

\begin{eqnarray}
W_Y&=&
\epsilon_{ijklm}\left(10_a^{ij}f_{ab}10_b^{kl}\Phi^m+
10_a^{ij}f_{1ab}10_b^{kl}{\Sigma^m_n\over M_{Pl}}\Phi^n+
10_a^{ij}f_{2ab}10_b^{kn}\Phi^l{\Sigma^m_n\over M_{Pl}}\right)\nonumber\\
&+&\bar\Phi_i10_a^{ij}g_{ab}\bar 5_{bj}
+\bar\Phi_i{\Sigma^i_j\over M_{Pl}}10_a^{jk}g_{1ab}\bar 5_{bk}
+\bar\Phi_i10_a^{ij}g_{2ab}{\Sigma_j^k\over M_{Pl}}\bar 5_{bk}\;,
\end{eqnarray}

\noindent
where $10$ and $\bar 5$ are the fermionic supermultiplet 
representations. After taking the SU(5) vev $\langle\Sigma\rangle=
\sigma\;diag(2,2,2,-3,-3)$ we get at $M_{GUT}$

\begin{eqnarray}
\label{alleq}
Y_U&=&4\left(f+f^T\right)
-12{\sigma\over M_{Pl}}\left(f_1+f_1^T\right)
-2{\sigma\over M_{Pl}}\left(4f_2-f_2^T\right)\;,\nonumber\\
\underline A&=&4\left(f+f^T\right)
+8{\sigma\over M_{Pl}}\left(f_1+f_1^T\right)
+2{\sigma\over M_{Pl}}\left(f_2+f_2^T\right)\;,\nonumber\\
\underline B&=&4\left(f+f^T\right)
+8{\sigma\over M_{Pl}}\left(f_1+f_1^T\right)
+4{\sigma\over M_{Pl}}\left(3f_2-2f_2^T\right)\;,\nonumber\\
Y_D&=&-g
+3{\sigma\over M_{Pl}}g_1
-2{\sigma\over M_{Pl}}g_2\;,\nonumber\\
Y_E&=&-g
+3{\sigma\over M_{Pl}}g_1
+3{\sigma\over M_{Pl}}g_2\;,\nonumber\\
\underline C&=&-g
-2{\sigma\over M_{Pl}}g_1
+3{\sigma\over M_{Pl}}g_2\;,\nonumber\\
\underline D&=&-g
-2{\sigma\over M_{Pl}}g_1
-2{\sigma\over M_{Pl}}g_2\;.
\end{eqnarray}

In the limit $M_{Pl}\to\infty$ we recover the old relations, 
but for finite $\sigma/M_{Pl}\approx 10^{-3}-10^{-2}$ one 
can correct the relations between Yukawas and at the same time 
have some freedom for the couplings to the heavy triplets. 

Clearly, due to SU(5) breaking through $\langle\Sigma\rangle$, 
the $T$, $\bar T$ couplings are different from the $H$, $\bar H$ 
couplings. However, under the SU(4) symmetry discussed before 
$\underline A\leftrightarrow\underline B$, 
$\underline C\leftrightarrow\underline D$ and 
$Y_U\leftrightarrow Y_U^T$. Only the terms that probe 
$\langle\Sigma_\alpha^\alpha\rangle -\langle\Sigma_4^4\rangle$ 
can spoil that; this is why $f_1$ and $g_1$ still keep $Y_U=Y_U^T$, 
$\underline A=\underline B$ and $\underline C=\underline D$. 

Notice that worrying about proton decay in a theory with 
$Y_D=Y_E$ may not be reasonable. Namely, suppose that you 
wish to preserve this relation; you can do that easily in 
eq. (\ref{alleq}) if you take $g_2=0$. It is amusing that 
you can set $\underline C=\underline D=0$ by choosing 
$g+2(\sigma/M_{Pl})g_1=0$ (for the third generation this 
requires a not too large $\tan{\beta}$ and/or $M_{Pl}$, 
which could be $M_{string}$, below $10^{18}$ GeV). In this way 
you can simply decouple the heavy triplets $T$ and $\bar T$ 
(an old idea of Dvali \cite{Dvali:1992hc}) 
and get rid of the $d=5$ proton 
decay. Obviously we do not wish to advocate this. After all, 
the idea of introducing higher dimensional operators, while 
preserving the SU(5) minimality structure all the way to $M_{Pl}$, 
is precisely to avoid $Y_D=Y_E$ for the first two generations. 
Notice further that with $Y_D\ne Y_E$ you can not set both 
$\underline C$ and $\underline D$ to vanish. You may be tempted to 
make $\underline A$ and $\underline B$ vanish, but that can not 
work for the large top Yukawa coupling. 

Although for realistic Yukawas the matrices $\underline A$, 
$\underline B$, $\underline C$ and $\underline D$ are not 
completely arbitrary, there is a new freedom not present if 
(\ref{abcdmin}) are valid: the fermion mixing matrices defined 
by (\ref{defx}) are no more related by (\ref{xmin}), but can 
be chosen freely, as long as (\ref{vckm}) are satisfied. This 
freedom could further diminish the decay amplitude.

In short, correct mass relations add more uncertainty to the 
proton decay amplitude and it may be worthwhile to perform a more 
quantitative analysis. 

\section{Sfermion and fermion masses and mixings and their 
impact on ${\bf \tau_p}$}

Last, but not least, the $d=5$ proton decay is generated through 
Yukawa couplings, and thus fermion and sfermion masses and 
mixings play an important role. Unfortunately, we know nothing about 
supersymmetry breaking, except that we must satisfy the 
experimental limits on FCNC. Ideally we wish to keep all the 
sparticles below TeV, but then FCNC becomes a serious issue, although 
still under control for carefully chosen mixings. Furthermore, if 
we ignore both (i) and (ii) the limits from proton decay can be 
used to rule out the minimal SU(5) theory. In all honesty, we do not 
know what that means, for this theory is obviously already ruled out 
by the wrong fermion relations. 

Now, the FCNC are mainly a problem for the first two generations; 
a popular approach is to assume the first two generations of 
sfermions heavy ($m\approx 10$ TeV), the so called decoupling 
regime \cite{Pomarol:1995xc,Dvali:1996rj,Cohen:1996vb}. 
In this case it is enough that the third generation of 
sfermions does not have large mixings with {\it both} of the first 
two generations of sfermions.

One can also worry about the naturalness 
\cite{Ellis:1986yg,Barbieri:1987fn,Dimopoulos:1995mi,Feng:1999zg}. 
Through the large top 
Yukawa couplings, the formula (\ref{higgs}) becomes here 
($i=1,2,3$) (for large $\tan{\beta}$ there are similar 
contributions of (s)bottom and (s)tau)

\begin{equation}
\label{higgsu}
m_h^2\approx m_0^2+{y_t^2\over 16\pi^2}\left[
(\tilde U^\dagger U)_{i3}(U^\dagger\tilde U)_{3i}\tilde m_i^2+
(\tilde U_c^\dagger U_c)_{i3}(U_c^\dagger\tilde U_c)_{3i}
\tilde m_i^{c2}\right]\;,
\end{equation}

\noindent
where $\tilde m_i$ and $\tilde m_i^c$ are left-handed and 
right-handed squark masses. Here and in the following we ignore 
the left-right mixing proportional to the small ratio 
$M_Z/m_{3/2}$; in fact, as long as $\tan{\beta}>10$, 
the LR mixing can be safely put even to zero without contradicting 
the experimental constraints on the Higgs mass \cite{Carena:2002es}. 

Now, for $\tilde m_a\approx\tilde m_a^c\approx 10$ TeV ($a=1,2$) 
in the decoupling regime, large $(\tilde U^\dagger U)_{a3}$ or 
$(\tilde U_c^\dagger U_c)_{a3}$ would imply a small amount of 
fine-tuning ($\approx 1\%$) in (\ref{higgsu}). Hereafter, we accept
that. No fine-tuning whatsoever, although appealing, to us seems 
exaggerated; after all it would eliminate large extra dimensions as 
a solution to the hierarchy problem. 

Strictly speaking, one could then ask why not simply push $\tilde m_3$ 
and $\tilde m_3^c$ all the way up to $10$ TeV and be safe? A sensible 
point, but as we said before, we wish to have as many as possible 
superpartners below TeV and thus hopefully detectable at LHC. 
In other words, all the gauginos and Higgsinos and the third generation 
of sfermions are assumed to have masses lower or equal TeV, we only 
take $\tilde m_{1,2}\approx 10$ TeV or so. 

In this case, we need to worry only about the third generation of 
sfermions. We also assume light gauginos and Higgsinos, 
$m\approx 100$ GeV. We have recently performed a detailed 
analysis of all $d=5$ proton decay amplitudes neglecting the mixing 
of left and right sfermions \cite{Bajc:2002bv}. An interesting 
question is: can the contribution of the third generation of 
sfermions be made arbitrary small? Remarkably enough, the answer 
is yes, i.e. it can be set even to zero, or, even easier, it can 
be small enough to keep $\tau_p$ above the experimental limit. 

The solution is the following. In our paper \cite{Bajc:2002bv} 
we give a typical set of constraints need to make the proton 
decay small ($a,b=1,2$):

\begin{eqnarray}
\label{constr}
&&(\tilde U^\dagger D)_{3a}\approx 0\;\;\;,\;\;\;
(\tilde D^\dagger D)_{3a}\approx 0\;\;\;,\;\;\;
(\tilde E_c^\dagger E_c)_{3a}\approx 0\;\;\;,\nonumber\\
&&(\tilde N^\dagger E)_{3a}\approx 0\;\;\;,\;\;\;
(\tilde D_c^\dagger D_c)_{3a}\approx 0\;\;\;,\;\;\;
(\tilde E^\dagger E)_{3a}\approx 0\;\;\;,\nonumber\\
&&(\tilde U_c^T Y_U^TD)_{3a}\approx 0\;\;\;,
\end{eqnarray}

\noindent
and

\begin{equation}
\label{groza}
A_0=\epsilon_{ab}(D^T\underline C\tilde N)_{a3}
(\tilde U^T\underline AD)_{3b}\approx 0\;\;\;.
\end{equation}

The constraints (\ref{constr}) can clearly be satisfied exactly
by the sfermion mixing matrices at $1$ GeV. It is reassuring that 
the sfermionic sector does not break strongly SU(2). This is 
consistent with the SU(2) invariance of the soft masses, which 
dominate the total sfermion masses. The last constraint, 
eq. (\ref{groza}), can be satisfied in the approximation 
$\underline C=Y_D=Y_E$, which is true in the minimal 
renormalizable model, but at $M_{GUT}$, not at 1 GeV. The 
relation $\underline C=Y_D=Y_E$ is however not stable under 
running. To get an idea of how big this contribution is 
at the electroweak scale, one can take the approximation that 
the Yukawas do not run. In the leading order in small Yukawas 
(except for $y_t$) one gets 

\begin{equation}
A_0\approx y_cy_\tau V_{33}^*V_{23}V_{32}V_{21}
\left[1-\left(M_Z/M_{GUT}\right)^{y_t^2/16\pi^2}\right]
x_1^{-1/33}x_2^{-3}x_3^{4/3}\;,
\end{equation}

\noindent
where $V$ is the CKM matrix and $x_i=\alpha_i(M_Z)/\alpha_U$. 
There is only one non-vanishing diagram (the rest vanishes due 
to (\ref{constr})) and it 
is proportional to $V_{13}A_0$: fortunately, this seems to be 
small enough. On top of this, in the amplitude the combination 
(\ref{groza}) gets multiplied with a combination of neutralino 
soft masses $m_{\tilde w_3}$ and $m_{\tilde b}$, which can be 
fine-tuned to an arbitrary small (or even zero) value. 
And, of course, we must keep in mind (i) and (ii), which 
tell us that $m_T$ can be large and $\underline A$ and/or 
$\underline C$ completely different than $Y_U$ ($Y_E$). 

\section{Constraints from FCNC}
Although there is no realistic theory of sfermion soft terms, 
there are low energy constraints on sfermion mixings. These 
come from the flavour changing neutral currents phenomena: 
$\mu\to e\gamma$, $b\to s\gamma$, $B-\bar B$, $K-\bar K$, etc. 
It is easy to see, that the combinations which appear in 
(\ref{constr}) are exactly the ones that appear in these 
flavour changing processes. So they automatically take care also 
of these low-energy experimental data. The only flavour changing 
processes that could get sizeable contributions are the ones 
which involve up type sfermions like for example $D-\bar D$ or 
$c\to u\gamma$. These are not constrained by (\ref{constr}), but 
at the same time are not very much constrained by the low-energy 
experiments, so they do not represent a real issue at this stage. 

Constraints (\ref{constr}) are not unique. One can find other 
relations between sfermion and fermion mixing matrices that make 
the proton decay amplitude zero or small. However, typically, these 
solutions can be dangerous for FCNC processes, since they do 
not au\-to\-ma\-ti\-cally cancel their contributions. So one has to 
analyze the FCNC processes case by case. At the present day status 
(or ignorance) of proton decay and FCNC experiments we believe 
that this is premature. 

\section{Summary}
We have emphasized here three major sources of uncertainties in 
estimating the $d=5$ proton decay. These are, in no particular order: 
(i) the ignorance of the masses of the color octet and weak triplet 
supermultiplets in the adjoint Higgs; this can easily increase the 
proton lifetime by a factor of thousand or so; (ii) higher dimensional 
operator correction of fermion masses; more difficult to quantify, but 
not necessarily less important; (iii) the ignorance of sparticle 
masses and mixings; although somewhat artificial, this possibility 
alone is enough to keep $\tau_p>(\tau_p)_{exp}$. 

In short, there is at least $10^3$ uncertainty in predicting $\tau_p$, 
and possibly as large as $10^4$ or bigger. Since none of the points 
(i), (ii), (iii) requires any change of the structure of the theory, 
the minimal supersymmetric SU(5) GUT is still in accord with all the 
experimental constraints. It is true, though, that the parameter space 
is becoming small and improvement in $(\tau_p)_{exp}$ is badly needed.

\section*{Acknowledgments}

B.B. thanks the organizers for the interesting and stimulating 
conference. The work of B.B. is supported by the Ministry of 
Education, Science and Sport of the Republic of Slovenia; the work 
of G.S. is partially supported  by EEC under the TMR contracts 
ERBFMRX-CT960090 and HPRN-CT-2000-00152. Both B.B. and P.F.P thank 
ICTP for hospitality during the course of this work.

\end{document}